\documentclass[twocolumn]{aastex631}
\usepackage{amsmath}
\usepackage[none]{hyphenat}
\usepackage{enumitem}
\setlist{noitemsep}

\usepackage{wasysym}
\let\tablenum\relax 

\usepackage{hyperref}
\begin{document}
\title{LUMOS : Linear programming Utility for Multi-messenger Optical Scheduling}

\author[0000-0002-5890-9298]{Yogesh P. Wagh}
\affiliation{Astrophysics Research Institute, Liverpool John Moores University, 146 Brownlow Hill, Liverpool, L3 5RF, UK}
\affiliation{Department of Physics, Indian Institute of Technology Bombay, Powai, 400 076, India}

\author[0000-0002-8262-2924]{Michael W. Coughlin}
\affiliation{School of Physics and Astronomy, University of Minnesota, Minneapolis, Minnesota 55455, USA}

\author[0000-0001-9898-5597]{Leo P. Singer}
\affiliation{Astrophysics Science Division, NASA Goddard Space Flight Center, Code 661, Greenbelt, MD 20771, USA}
\affiliation{Joint Space-Science Institute, University of Maryland, College Park, MD 20742, USA}

\author[0000-0002-6112-7609]{Varun Bhalerao}
\affiliation{Department of Physics, Indian Institute of Technology Bombay, Powai, 400 076, India}

\begin{abstract}
The detection of GW events by LIGO-Virgo-KAGRA has opened new avenues for multi-messenger astrophysics; however, electromagnetic counterparts remain elusive due to large localization uncertainties. Wide-field optical surveys like the Zwicky Transient Facility (ZTF) play a crucial role in follow-up, but efficient scheduling is essential. In this work, we present \texttt{LUMOS}, a Mixed Integer Linear Programming (MILP) approach that selects fields via a maximum coverage problem and schedules observations to maximize cumulative probability while respecting observability constraints. Using 1199 GW events from O4, we compare \texttt{LUMOS} scheduler with \texttt{gwemopt}, showing an 84.7\% higher mean cumulative probability and better performance in nearly all cases. While designed for ZTF, \texttt{LUMOS}'s framework parallels the M\textsuperscript{4}OPT toolkit for space missions, highlighting the broader applicability of MILP-based scheduling to both ground- and space-based follow-up.

\end{abstract}


\section{Introduction} \label{sec:intro}
Gravitational-wave (GW) astronomy has revolutionized our understanding of the universe, providing unprecedented insights into astrophysical processes. The observation of GW170817 
 \citep{Abbott_2017_gw} in August 2017 marked the beginning of a new era in multi-messenger astronomy, leading to groundbreaking discoveries in $r$-process nucleosynthesis \citep{Hotokezaka_2018,Chornock_2017}, constraints on the neutron star equation of state \citep{Neutron_star_eos,Radice_2018,10.1093/mnrasl/slz133,doi:10.1126/science.abb4317} and new estimates of the Hubble constant \citep{2017_hubble,doi:10.1126/science.abb4317,hotokezaka2018hubbleconstantmeasurementsuperluminal}. This event was accompanied by electromagnetic counterparts \citep{2017ApJ...848L..12A,doi:10.1126/science.aap9455,doi:10.1126/science.aap9580,doi:10.1126/science.aaq0073,Pian_2017,doi:10.1126/science.aaq0186,Smartt_2017}, including a kilonova, a short gamma-ray burst \citep{Goldstein_2017}, and an afterglow \citep{doi:10.1126/science.aap9855,Troja_2017}. The alert triggered extensive follow-up observations, leading to the identification of the optical counterpart, AT2017gfo \citep{Coulter_2017}. 
 However, despite multiple neutron star merger detections in subsequent LIGO-Virgo-KAGRA (LVK) observing runs \citep{PhysRevX.11.021053}, including the recent O4a catalog release with 128 new compact binary candidates \citep{bns128} and the public O4a dataset, confirmed electromagnetic counterparts have remained elusive. Many events have been located beyond the range of telescopes, while others have highly uncertain localizations \citep{10.1093/mnras/staa1925,Petrov_2022}, making targeted electromagnetic follow-up challenging. As detections accumulate, improving follow-up strategies is critical to maximizing scientific yield and advancing our understanding of compact object mergers, cosmic expansion, and fundamental physics.

ZTF is a time-domain survey that uses the wide-field 48-inch Samuel Oschin Telescope, at the Palomar observatory, with large format cameras giving it a field of view (FOV) of 47 square degrees \citep{Bellm_2018,Masci_2018,Graham_2019,Dekany_2020}. Its primary science goals are to study cosmic explosions, probe multi-messenger astrophysics, and investigate variable stars and Solar System objects. ZTF plays a crucial role in optical follow-up observations of GW sources, where its large FOV provides a significant advantage. Rapid response is essential for important gravitational wave (GW) or neutrino events, requiring quick triggering of observatories to facilitate follow-up observations. Scheduling plays a key role in this process, determining how observations are assigned in response to multi-messenger events. ZTF operates with a fixed set of 1,778 fields covering the region of the sky accessible to it,
split equally between primary and secondary fields \citep{Bellm_2018}. Secondary fields help compensate for chip gaps in primary fields, ensuring comprehensive coverage. 



Transients like GW events detected by the LVK, often have large localization uncertainties exceeding a thousand square degrees \citep{article}, requiring observations of a large number of fields. Each field is imaged at least twice with a gap of at least 30~min \citep{bellm2014zwickytransientfacility}. This helps filter out transient contaminants, such as asteroids, which can move noticeably within that time frame. Sometimes, the second visit may be in a different filter to obtain multi-filter photometry for colour characterization of astrophysical transients. Given the sky localisation of a transient, a scheduling algorithm determines which fields  to observe and in what order. A straightforward approach to scheduling is to maximize the probability coverage of the probability skymap\footnote{ligo.skymap - Python Module: \url{https://lscsoft.docs.ligo.org/ligo.skymap/}} across all selected fields, which serves as the objective function \citep{10.1093/mnras/sty1066}. 

Current scheduling approaches often rely on computationally intensive greedy algorithms, which follow a problem-solving heuristic that selects the locally optimal choice at each step. This method is used by the \texttt{gwemopt} scheduler \citep{Coughlin_2019,10.1093/mnras/staa1498} and \texttt{Astroplan} \citep{Morris_2018}. However, greedy algorithms do not produce globally optimized solutions, which can be advantageous for improving scheduling efficiency. Since greedy approaches lack foresight, they fail to plan observations across all time domains—beginning, middle, and end of an observing window—limiting overall efficiency. The scheduling problem is combinatorially complex and challenging to solve. A viable approach to making these problems computationally manageable is Integer Linear Programming (ILP). ILP allows problem formulation with discrete integer variables, linear constraints, and an objective function that can be maximized or minimized. Several observatories also use MILP-based schedulers to optimize their observations. For example, the Las Cumbres Observatory (LCO) employs an ILP approach to schedule observations across its network of identical imagers and spectrographs, aiming to maximize the number of completed observations while considering priorities set by the Time Allocation Committee \citep{lampoudi2015integerlinearprogrammingsolution}. Similarly, ALMA utilizes an ILP model to optimize scheduling by balancing scientific priorities, program completion, and telescope efficiency \citep{SOLAR201690}. ZTF implemented a MILP scheduling approach that optimized nightly observations by assigning targets to temporal blocks, minimizing slew times and filter changes, while incorporating time-varying observational conditions to maximize transient discovery efficiency \citep{2019PASP..131f8003B}. In this paper, we leverage ILP to solve a maximum coverage problem and employ a mixed-integer linear programming (MILP) approach to develop an optimized scheduler. Mixed ILP incorporates continuous variables alongside integer variables, enhancing flexibility and optimization. MILP provides an efficient and globally optimized approach to scheduling follow-ups. Unlike greedy algorithms, like the one used in \texttt{gwemopt}, MILP incorporates foresight into observation planning, ensuring that selected fields are observed at optimal times.

In this paper, we introduce \texttt{LUMOS}, an MILP-based scheduler for multi-messenger follow-up using a wide-field survey telescope like ZTF. \texttt{LUMOS}, which stands for Linear programming Utility for Multi-messenger Optical Scheduling, leverages Mixed Integer Linear Programming to optimize field selection and observation timing. The problem setup is outlined first in Section~\ref{sec:PROBLEM SETUP}, describing the observability constraints that influence which GW events and sky regions can be targeted for follow-up. These constraints include localization uncertainty of GW events and field visibility throughout the night. The scheduler's strategy for selecting the optimal set of fields to maximize the probability of detecting an electromagnetic counterpart is then explained. Next, the scheduling algorithm is presented in Section~\ref{sec:MILP}, detailing the MILP formulation used to optimize the observation plan. Key variables, constraints, and the objective function that guide the scheduling process are defined. The algorithm accounts for telescope slewing times, non-overlapping observations, cadence constraints, and visit sequencing to ensure efficient coverage of target regions. To evaluate \texttt{LUMOS}, its performance is compared against \texttt{gwemopt}, a widely used scheduling method. In Section~\ref{sec:results}, we consider all GW events published by the LVK during O4 across all search pipelines—including candidates consistent with terrestrial noise, binary black hole (BBH), binary neutron star (BNS), and neutron star–black hole (NSBH) mergers—without applying any significance threshold. Probability coverage, field selection efficiency, and overall scheduling effectiveness are then assessed. Results demonstrate that \texttt{LUMOS} achieves a significant improvement in cumulative probability coverage over \texttt{gwemopt} while maintaining efficient field selection. The implications of these findings, potential improvements to the scheduler, and future directions for optimizing multi-messenger follow-up strategies with wide-field surveys are discussed.

\section{Localization Coverage Strategy} \label{sec:PROBLEM SETUP}

ZTF observes specific sky regions by dividing the entire sky into predefined fields. GW events detected by the LVK provide skymaps, represented as probability distributions in HEALPix format \citep{Singer_2016}. HEALPix (Hierarchical Equal Area isoLatitude Pixelization) is a mapping scheme that divides the celestial sphere into equal-area pixels, enabling efficient representation of sky data \citep{Gorski_2005}. It is commonly used for astrophysical skymaps, such as GW localizations, where each pixel contains a probability value indicating the event's likely origin. To effectively follow up on these events, observations are prioritized within the 90\% confidence region of the skymap, ensuring coverage of the most probable locations.


The start time of observations is determined by when the event occurs. If the event happens during the night, observations begin right away. If it occurs during the day, observations start at sunset, using astronomical twilight as the reference and based on the observer’s location. 

Each ZTF field is treated as a separate target. The visibility window for each field starts when it rises above the horizon and ends when it sets. Circumpolar fields, which remain above the horizon, are treated as always observable, with their visibility ending when observations conclude. After filtering, each field is assigned a well-defined start and end time.

Observability constraints are applied to all selected fields. Since electromagnetic counterparts to GW events require multiple observations in different filters, fields must remain observable for multiple exposures.  ZTF follows up transients using a 30-minute cadence while alternating between filters such as r$^\prime$, i$^\prime$ and g$^\prime$. This strategy helps distinguish between transient sources like asteroids while providing color and magnitude evolution information.

ZTF detects numerous transients, and monitoring magnitude evolution enables the construction of light curves, which helps eliminate irrelevant transients. New sources are identified through nightly image differencing against archival references, with alerts passed through a machine-learning filter and crossmatched with catalogues to exclude artifacts and known variables \citep{Andreoni_2021}. Refined photometry and light curve fitting then allow the discovery of genuine fast transients, which can be rapidly prioritized for follow-up. The selected fields must be visible during the dark period between evening and morning astronomical twilight, while ensuring they remain above an airmass of 2.5 during their visibility window. These fields are further filtered based on the probability they cover. Fields with an individual probability contribution of less than 0.0001 are excluded from scheduling. This threshold ensures a manageable number of fields for observation while discarding those with negligible contributions to the total localization probability. Including such low-probability fields would have minimal impact on the overall coverage and would unnecessarily complicate the scheduling process.

\subsection{Mathematical Formulation}
In case the fields have zero overlap with each other, this problem becomes trivial. However, in practice, certain parts of the sky are present in multiple fields and the selection should be done while ensuring that the probabilities for those are counted only once.  
To systematically select the fields covering the event localization, we model this as an optimization problem. The objective is to maximize the probability coverage within the observational constraints. The problem is formulated as a Mixed Integer Linear Programming (MILP) model, which is modification of maximum weighted coverage problem \citep{Nemhauser}. 




Consider a pixel \( j \) with probability \( p_j \), covered by multiple fields that form the set \( S_j \). Let \( x_i \) be a binary variable indicating whether field \( i \) is selected, and let \( y_j \) be a binary variable representing whether pixel \( j \) in the skymap is selected. In other words, \( x_i \) and \( y_i \) are flags indicating the selection of fields and pixels respectively. The formulation is:
\begin{align}
    & \mathrm{maximize} \sum p_j y_j, \label{eq:obj_fun}\\
    & y_j = x_1 \lor x_2 \lor \dots  \quad \forall x \in S_j \label{eq:binary_sum}\\
    & \sum_{i} x_i \leq k .\label{eq:number_fields}
\end{align}
Equation~\ref{eq:obj_fun} defines the objective function, where we maximize the total summed probability across the selected pixels. In Equation~\ref{eq:binary_sum}, we calculate the logical \texttt{or} of the selection flags of all fields containing the pixel \( j \) to determine if the pixel was selected. Equation~\ref{eq:number_fields} applies the constraint that the total number of selected fields does not exceed \( k \), which is determined by the exposure time and number of visits possible within the night.

\begin{figure}[ht]
    \centering
    \includegraphics[width=\columnwidth]{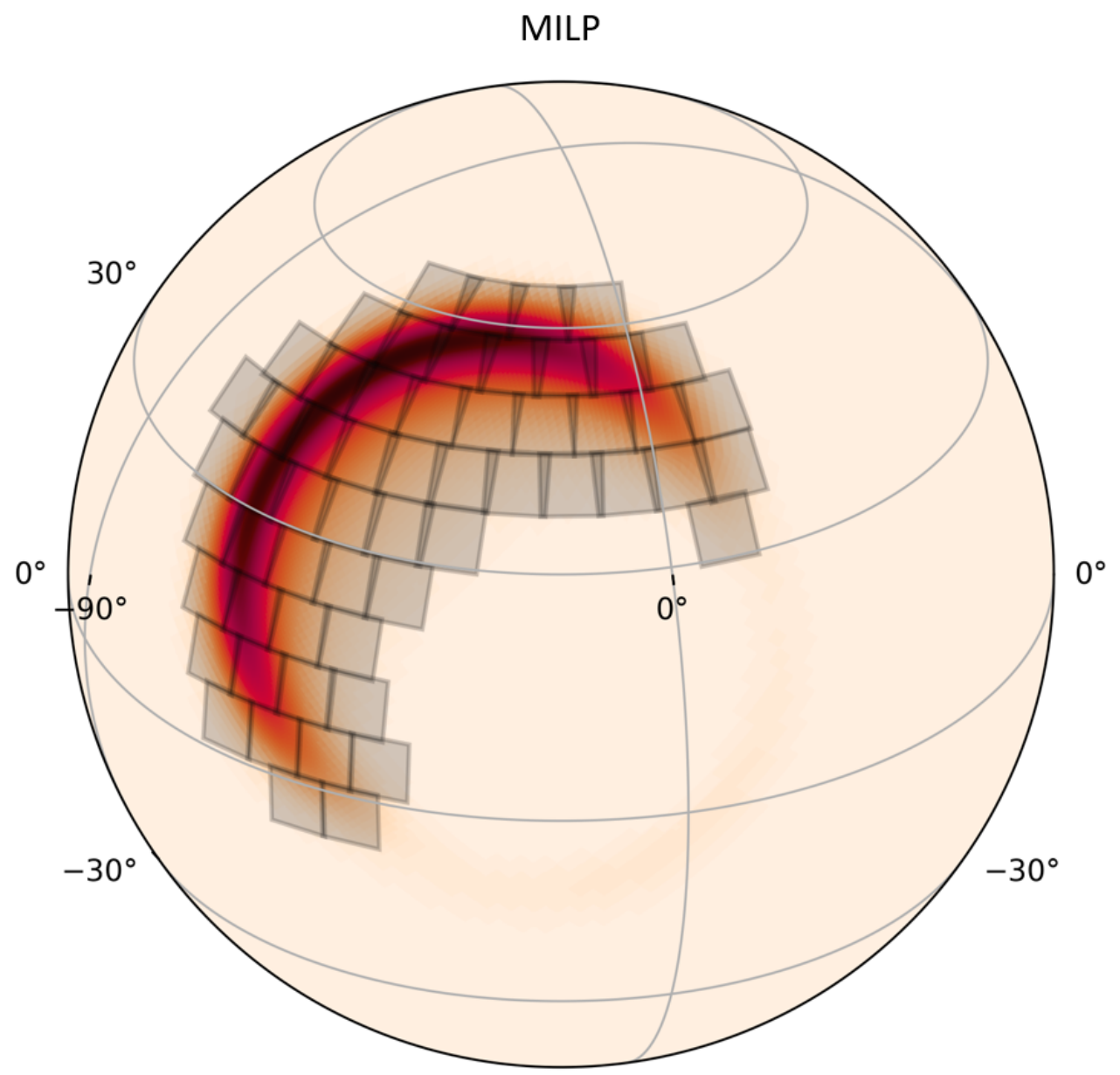}
    \caption{The highlighted tiles represent the fields covering the gravitational-wave event localization, obtained from the \texttt{LUMOS}'s MILP-based maximum coverage algorithm.}
    \label{fig:max cov pro}
\end{figure}

We use CPLEX by IBM\footnote{CPLEX Installation Guide: \url{https://web.archive.org/web/20250522160842/https://mertbakir.gitlab.io/operations-research/how-to-install-cplex-ibm-academic-initiative/}}$^,$\footnote{IBM ILOG CPLEX Product Page: \url{https://web.archive.org/web/20250522161058/https://www.ibm.com/products/ilog-cplex-optimization-studio}} as the MILP solver to optimize the selection of fields, ensuring that the highest probability regions are observed while adhering to the observational constraints. The selected fields from this process form the basis for the scheduling phase, where observation times are assigned to maximize transient detection efficiency. CPLEX is a widely used industry-grade optimization software designed for solving complex mathematical programming problems, including MILP. It provides efficient solvers for large-scale optimization tasks, making it suitable for applications in scheduling, logistics, and operations research. While commercially available, CPLEX offers a free student license for academic research and educational use.

At the end of this process, we obtain a set of fields (Figure~\ref{fig:max cov pro}) that can be used as an input for the scheduling step.


\section{Scheduling Algorithm} \label{sec:MILP}
A critical aspect of transient follow-up is distinguishing astrophysical sources from false detections. The 30-minute re-observation constraint plays a crucial role in this process. By enforcing a minimum time gap between successive observations of the same field, the scheduler ensures that short-lived moving objects, such as asteroids, are effectively ruled out. Additionally, multi-filter observations provide color information, aiding in classification. This strategy enhances the robustness of transient identification, reducing false positives and improving the efficiency of electromagnetic counterpart searches.

We now formalize the scheduling problem using a Mixed Integer Linear Programming (MILP) approach. The goal is to design an observation sequence that visits a set of pre-selected fields multiple times—three times, in our case—to maximize the total detection probability while satisfying temporal and operational constraints.

\begin{figure}[ht]
    \centering
    \includegraphics[width=1\linewidth]{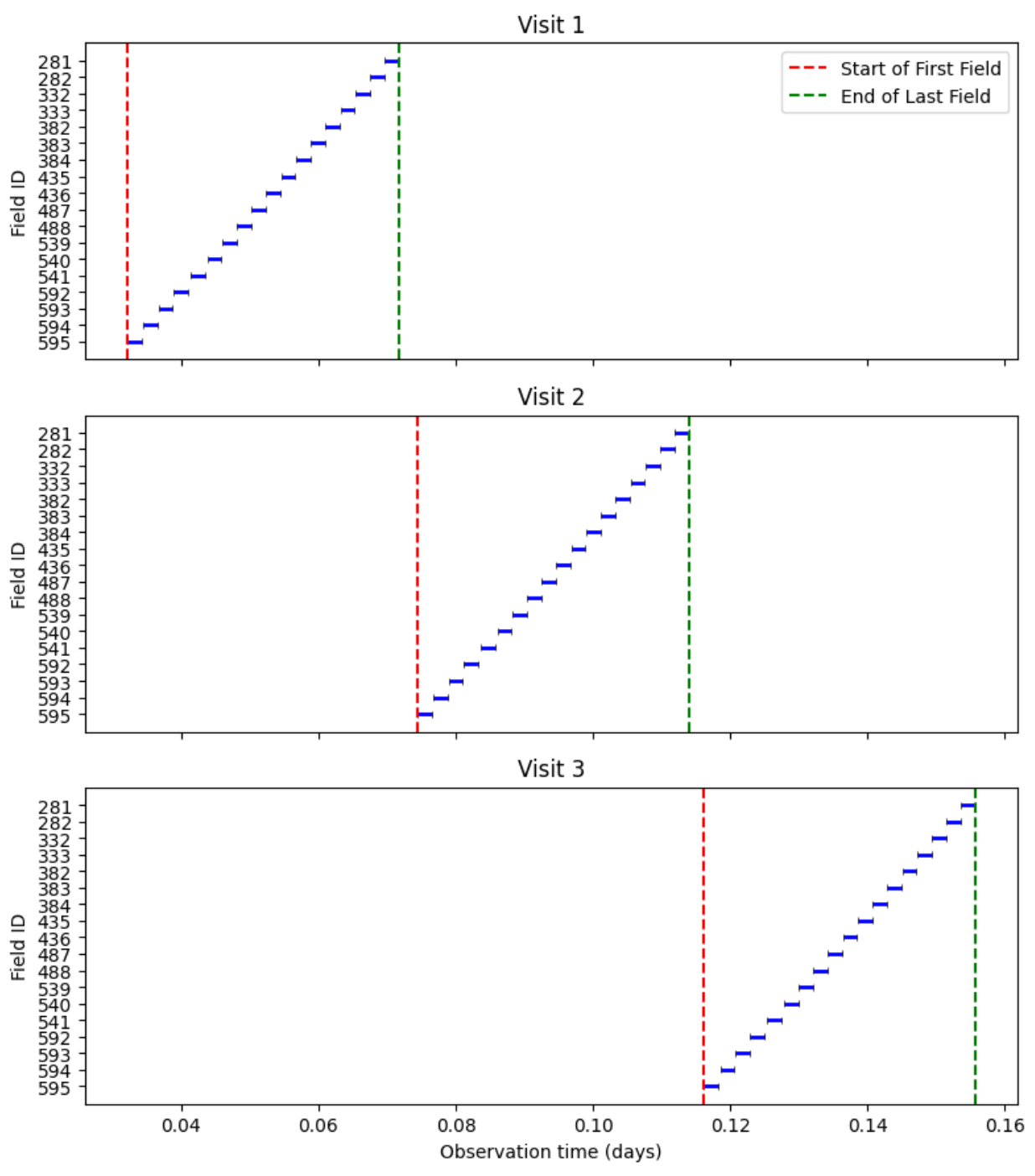}
    \caption{Illustration of the scheduling process for multiple visits.}
    \label{fig:multiple visit}
\end{figure}

Each visit is defined as a round in which the telescope observes every selected field once. The subsequent visit can begin only after the last field from the previous visit is completed (Figure~\ref{fig:multiple visit}). To model this, we define binary and continuous variables that capture whether a field is selected for a given visit, the start time of its observation, and a helper variable to separate visits. We also consider a precomputed slew time matrix that encodes the time required for the telescope to move from one field to another.

We will apply the Big-M method \citep{article_bigM}, a standard MILP technique that enables conditional constraints while preserving linearity. The variable \( t_{ij} \), which denotes the start time of observing field \(i\) during visit \(j\), is bounded between 0 and \( M \)—a large upper bound corresponding to the full observation window.

The selected fields from the solution to the maximum coverage problem are passed into the scheduler. The following variables and constants define our formulation:
\subsection{Variables}
\begin{enumerate}
    \item \( I = \{0, 1, \dots, n_r - 1\} \): Selected fields.\\
    \(n_r:\text{Number of fields}\)
    \item \( J = \{0, 1, \dots, n_v - 1\} \): Visit indices.\\
    \(n_v:\text{Number of visits}\)
    \item \( \Delta \): Exposure time for each field.
    \item \( \mathbf{M} \): A large constant representing the time window from observation start to end (Big-M).
    \item \( \mathbf{t}_{ij}, \quad i \in I, \quad j \in J \): Start time of observing field \(i\) in visit \(j\); bounded between 0 and \(M\).
    \item \( \mathbf{x}_{ij}, \quad i \in I, \quad j \in J \): Binary variable indicating whether field \(i\) is selected in visit \(j\).
    \item \( \mathbf{b}_{j}, \quad j \in J, j\geq1 \): A continuous helper variable representing the end time of visit \(j-1\), used to ensure that visit \(j\) starts only after the previous visit has fully concluded. It serves to isolate visits temporally by enforcing sequential scheduling.

    \item \( \mathbf{s}_{ik}, \quad i,k \in I\): Slew time between observing field \(i\) and field \(k\). Pre-computed using angular separation and telescope dynamics.
\end{enumerate}

Slew time is computed using the field separation \(d\) as:
\[
\text{s}(d) = 
\begin{cases}
    \sqrt{\frac{2d}{a_{\text{slew}}}}, & \text{if } d \leq \frac{v_{\text{slew}}^2}{a_{\text{slew}}} \\[8pt]
    \frac{2 v_{\text{slew}}}{a_{\text{slew}}} + \frac{d - \frac{v_{\text{slew}}^2}{a_{\text{slew}}}}{v_{\text{slew}}}, & \text{if } d > \frac{v_{\text{slew}}^2}{a_{\text{slew}}}
\end{cases}
\]
where:
\begin{itemize}
    \item \( d \): Angular separation between fields.
    \item \( v_{\text{slew}} \): Maximum slew speed.
    \item \( a_{\text{slew}} \): Slew acceleration.
\end{itemize}

\subsection{Constraints and Objective Function}
We define the following mathematical constraints and the objective function for the optimization problem.

\textbf{Cadence Constraint}:
    Ensure that field \(i\) is re-observed at least 30 minutes after its previous observation:
\begin{equation}
    t_{i,j} - t_{i,j-1} \geq c \cdot (x_{i,j} + x_{i,j-1} - 1)    
\end{equation}
    for all \(i \in I\), \(j \geq 1\), where \(c \) is the cadence time (minimum separation between successive observations of the same field).

\textbf{Non-overlapping Constraint}:
    For each visit, no two fields should be observed simultaneously:
    \begin{align}
    t_{i,j} + \Delta \cdot x_{i,j} + \text{s}_{i,k} - t_{k,j} &\leq M \cdot (2 - x_{i,j} - x_{k,j}) \\
    t_{k,j} + \Delta \cdot x_{k,j} + \text{s}_{i,k} - t_{i,j} &\leq M \cdot (-1 + x_{i,j} + x_{k,j})
    \end{align}

\textbf{Visit Separation Constraint}:
    Enforce separation between visits using a helper variable \(b\):
    \begin{align}
    t_{i,v-1} + \Delta \cdot x_{i,v-1} &\leq b_{v-1} \quad \text{(end of visit \(v-1\))} \\
    t_{i,v} &\geq b_{v-1} \quad \text{(start of visit \(v\))},
    \end{align}

    for \(v \geq 1\). 

\textbf{Objective Function}:
    Maximize cumulative probability of selected fields:

    \[
    \max \sum_{i=0}^{n_r - 1} \sum_{j=0}^{n_v - 1} \text{p}_i \cdot x_{i,j}
    \]

    where \(\text{p}_i\) is the detection probability for field \(i\).

A revisit scenario is illustrated in Figure~\ref{fig:multiple visit}. In this example, an exposure time of 3 minutes and a cadence interval of 30 minutes are used for scheduling. The blue horizontal segments represent exposures corresponding to the fields shown along the y-axis.

\section{Scheduling GW Events} \label{sec:results}

The detection of electromagnetic (EM) counterparts to GW events remains challenging due to large localization uncertainties and faint transient signals and their distance. Recent observing scenarios for LIGO-Virgo-KAGRA's O4 and O5 runs predict annual BNS detection rates of $36^{+49}_{-22}$ and $180^{+220}_{100}$, respectively, with NSBH rates significantly lower \citep{Kiendrebeogo_2023}. ZTF plays critical roles in follow-up, with predicted annual EM counterpart detections for BNS mergers ranging $0.43^{+0.58}_{-0.26}$ (ZTF) in O4. These projections emphasize the need for efficient scheduling to maximize coverage within limited observational windows.

\begin{figure*}[!htb]
    \centering
    \includegraphics[width=1\linewidth]{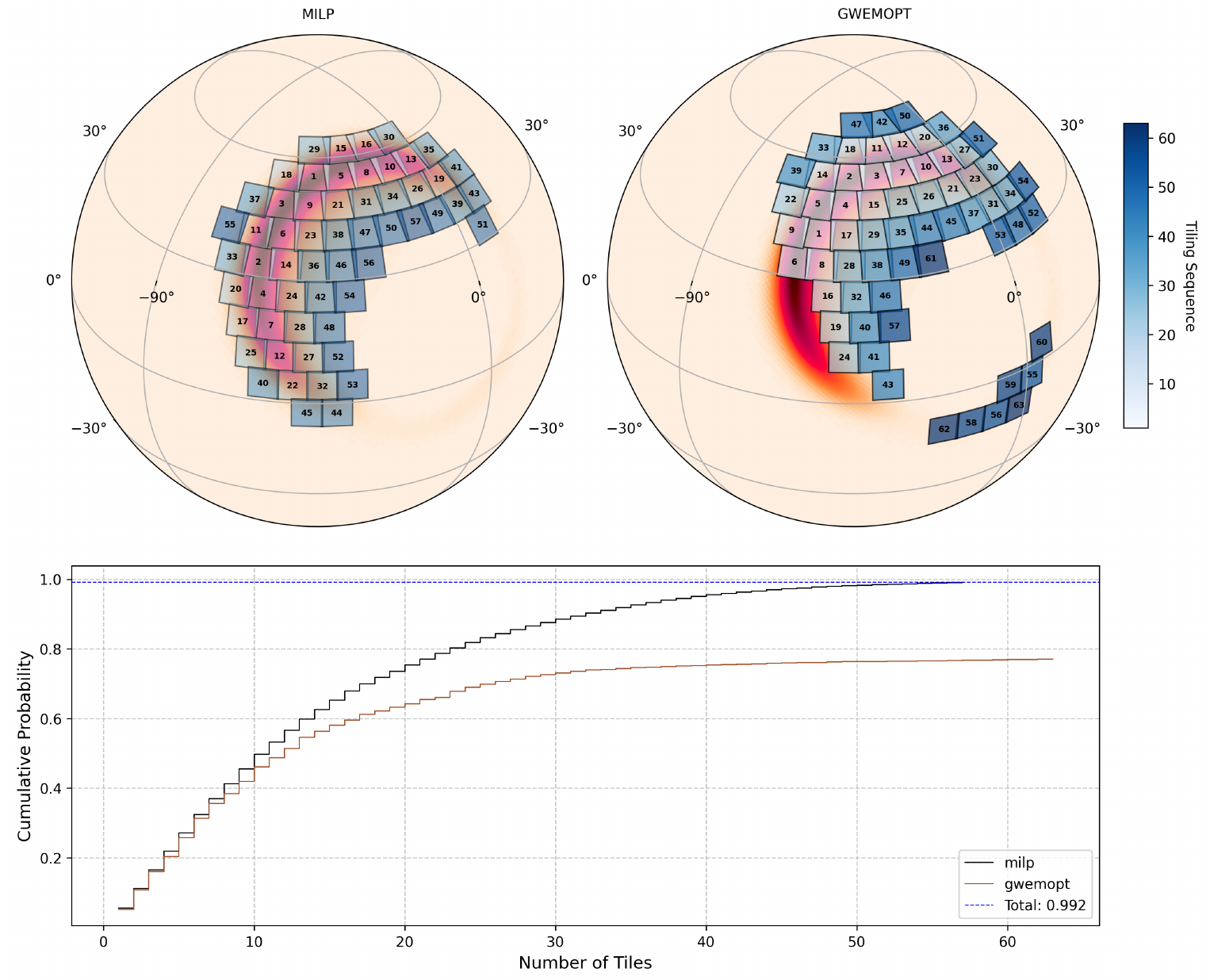}
    \caption{Comparing scheduling performance of \texttt{LUMOS} (upper left panel) and \texttt{gwemopt} (upper right panel). \texttt{LUMOS} correctly schedules observations of the entire high probability region, while \texttt{gwemopt} ends up missing a part of it, instead recommending observations of the low-probability part of the localisation. The bottom panel shows the cumulative probability coverage as a function of number of fields observed. \texttt{LUMOS} observations terminate faster with a higher cumulative probability coverage.}
    \label{fig:scheduled maps and cumulative probability}
\end{figure*}

Our analysis leverages 1199 GW events from O4\footnote{we select events from May 2023 to January 2025.} to benchmark the MILP scheduler against \texttt{gwemopt}. The \citet{Kiendrebeogo_2023} simulations emphasize that wide-field surveys must prioritize rapid cadence, multi-filter observations, and field revisits to distinguish kilonovae from contaminants, aligning with our 30-minute cadence constraint and slew-aware scheduling. Additionally, their predicted median 90\% credible sky areas of $1860^{+250}_{-170}$ deg$^2$ for BNS events in O4 highlight the necessity of algorithms that optimize tile selection under time-varying visibility.

The scheduling parameters were kept consistent across both algorithms. Each observation used an exposure time of 30 seconds, and each field was revisited three times to help distinguish genuine transient sources from asteroids. For ZTF ToO observations, longer exposures (e.g., 180–300 seconds) are also common when the localization region is sufficiently small, enabling deeper photometry. Our scheduler can readily accommodate such configurations as well. The observation start time is determined based on whether the event has risen above the horizon (see Section~\ref{sec:PROBLEM SETUP}). Once the start time is set, a follow-up plan is scheduled for the 12-hour window that follows --- to aid uniform comparison of the two algorithms. In practice, the observations may end earlier due to morning astronomical twilight, and that can easily be set as the end time in \texttt{LUMOS}.

Figure~\ref{fig:scheduled maps and cumulative probability} shows the cumulative probability covered by \texttt{gwemopt} and \texttt{LUMOS} for a representative event (superevent ID: G429012). MILP achieves higher cumulative probability with fewer fields, and this advantage becomes evident early in the observation. The numbers on each tile represent the sequence of field observations within a single visit. This schedule is repeated for the second and third visits with the same field order.

\begin{figure}[tbh]
    \centering
    \includegraphics[width=1\linewidth]{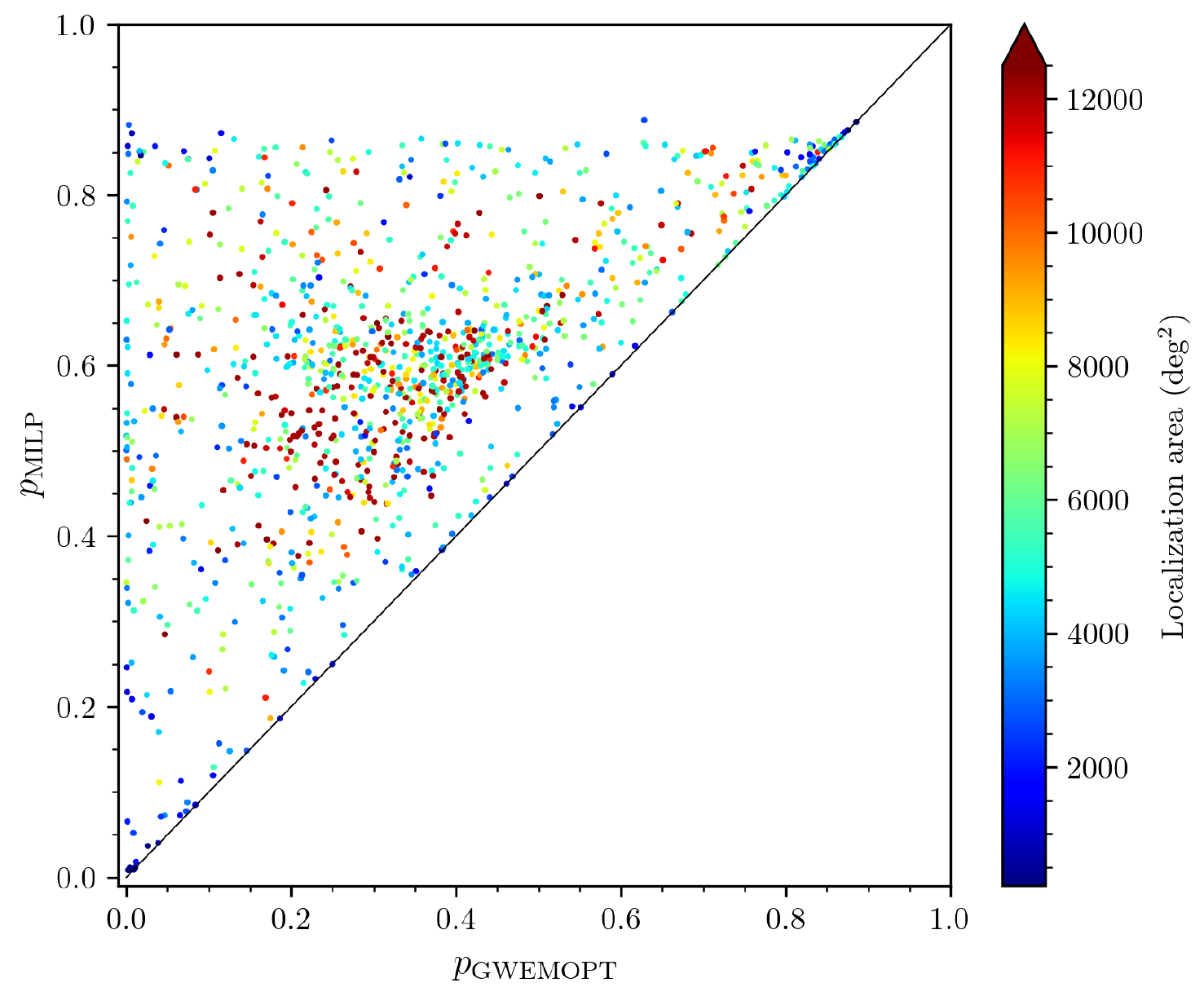}
    \caption{Comparison of cumulative probability coverage achieved by \texttt{LUMOS} and \texttt{gwemopt} for 1199 gravitational-wave events from O4. Each point represents a single skymap. The dashed diagonal line indicates equal performance. Points above the line correspond to cases where \texttt{LUMOS} outperforms \texttt{gwemopt}, which occurs in 99.7\% of cases. The color of each point reflects the sky area covered (in deg$^2$), providing additional context for the efficiency of coverage.}
    \label{fig:gwempot_v_milp_prob}
\end{figure}


The scheduling results are shown in Figure~\ref{fig:scheduled maps and cumulative probability}. We observe that \texttt{gwemopt} misses a part of the high-probability region within the localization, whereas \texttt{LUMOS} includes those fields in its observation plan. This leads to a significantly higher cumulative probability covered by \texttt{LUMOS}. Additionally, the plots indicate that fields with higher probability are prioritized and observed earlier in the schedule.  

Figure~\ref{fig:gwempot_v_milp_prob} compares the cumulative probabilities achieved by \texttt{gwemopt} and \texttt{LUMOS} across 1199 real events from O4. \texttt{LUMOS} outperforms \texttt{gwemopt} in 99.7\% of cases. In the small fraction of cases (0.3\%) where \texttt{gwemopt} achieves slightly higher coverage, the difference is negligible and results from a probability threshold of 0.0001 per field applied in \texttt{LUMOS}'s formulation. For 67 skymaps, \texttt{gwemopt} was unable to generate a schedule, while \texttt{LUMOS} successfully did so. These points appear close to the $y$ axis in Figure~\ref{fig:gwempot_v_milp_prob}.

\begin{table}[!th]
    \centering
    \caption{Comparison of \texttt{LUMOS} and \texttt{gwemopt} scheduling performance over our set of simulated events.}
    \begin{tabular}{lcc}
        \hline
        \textbf{Metric} & \textbf{\texttt{LUMOS}} & \textbf{\texttt{gwemopt}} \\
        \hline
        Mean Cumulative Probability & 0.59 & 0.32 \\
        Median Cumulative Probability & 0.61 & 0.31 \\
        Avg. Probability per Field & 0.0052 & 0.0049 \\
        Median Probability per Field & 0.0040 & 0.0036 \\
        \hline
    \end{tabular}
    \label{tab:stats}
\end{table}

Table~\ref{tab:stats} summarizes key metrics for the two algorithms. The average cumulative probability achieved by \texttt{LUMOS} for our simulated data set is 0.59, compared to 0.32 for \texttt{gwemopt}. The median values are 0.60 and 0.31, respectively, indicating that \texttt{LUMOS} improves the median probability coverage by approximately 70\%. While cumulative probability provides a direct measure of performance, an additional comparison is made using the average probability per field. On average, \texttt{LUMOS} achieves 0.0052 probability per field, compared to 0.0049 for \texttt{gwemopt}, with median values of 0.0040 and 0.0036, respectively.


\begin{figure}[!bht]
    \centering
    \includegraphics[width=1\linewidth]{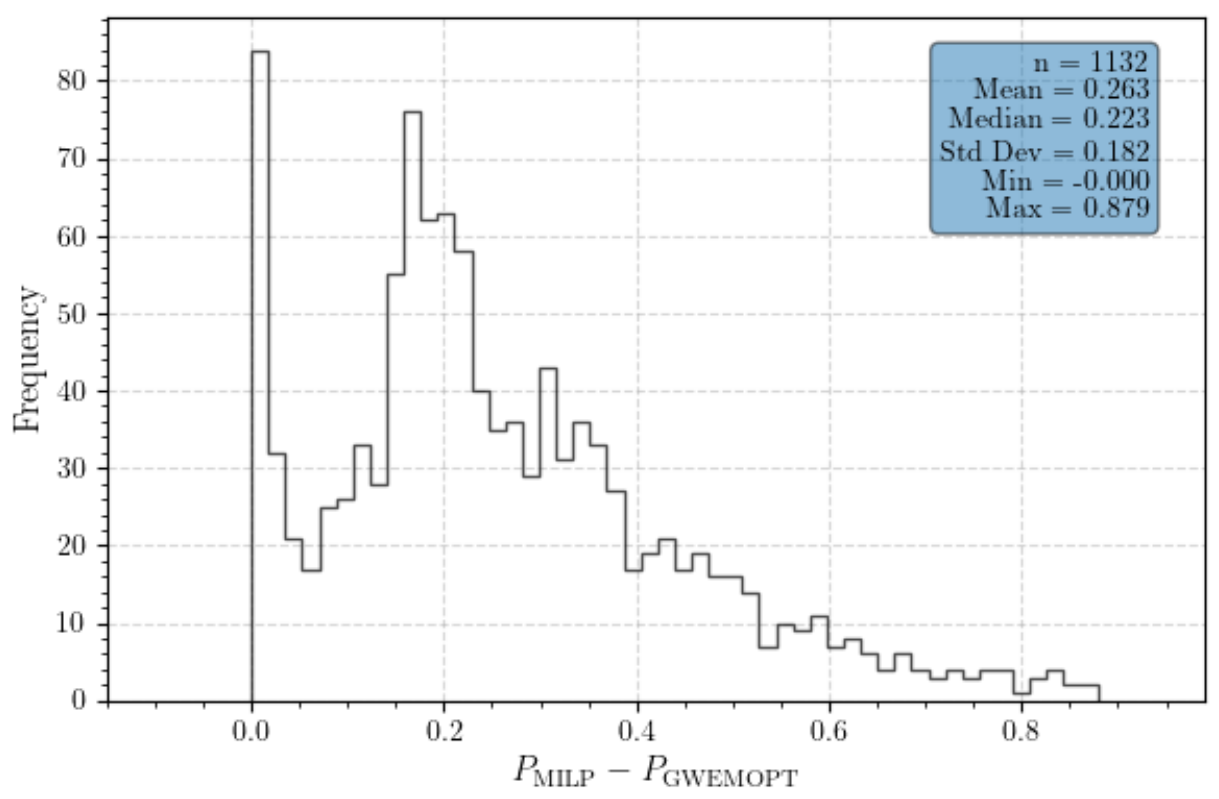}
    \caption{Histogram plot showing the difference of cumulative probability between \texttt{LUMOS} and \texttt{gwemopt}.}
    \label{fig:minus}
\end{figure}

Another way to compare the two algorithms is to examine the difference in cumulative probability covered by each, as shown in Figure~\ref{fig:minus}. We see that \texttt{LUMOS} outperforms \texttt{gwemopt}, providing better coverage and more efficient use of available fields, making it a more reliable choice for time-critical follow-up observations.


\section{Conclusion and Future Work}
In this paper, we introduce \texttt{LUMOS}, a scheduling algorithm for ground-based wide-field survey telescopes to optimize follow-up observations of multi-messenger events. We formulate the problem as a Mixed Integer Linear Programming (MILP) model, first solving the maximum coverage problem in Section~\ref{sec:PROBLEM SETUP} while incorporating observability constraints. We then present the scheduling algorithm in Section~\ref{sec:MILP} and compare its performance against the \texttt{gwemopt} scheduler.

To evaluate the effectiveness of \texttt{LUMOS}, we run both \texttt{gwemopt} and \texttt{LUMOS} on a set of O4 skymaps and analyze their cumulative probability coverage. The results, presented in Section~\ref{sec:results}, show that \texttt{LUMOS} achieves a significant increase in total probability coverage over \texttt{gwemopt}. Additionally, we compare the distribution of cumulative probabilities, demonstrating that \texttt{LUMOS} provides comparable or better probability coverage in all cases.

While the \texttt{LUMOS} scheduler improves upon existing methods, there are several areas for further development. The current formulation assumes a constant flux source, which is not accurate for rapidly evolving transients such as kilonovae. A natural improvement, already implemented in the \texttt{M\textsuperscript{4}OPT} framework \citep{Singer_2025} for the UVEX mission, is to incorporate light curve models directly into the scheduling process, allowing anticipated magnitude evolution to influence field selection and observation timing. \texttt{M\textsuperscript{4}OPT} also demonstrates additional features such as dynamic exposure-time optimization and foreground-dependent sensitivity variations, which could be adapted into our approach. Although our current scheduler does not yet include these capabilities, future work will focus on extending our model in this direction. Beyond adopting specific techniques, our results also suggest that lessons learned from our ground-based scheduling can inform the continued development of \texttt{M\textsuperscript{4}OPT}, bridging strategies between wide-field surveys like ZTF and space-based missions like UVEX.


Additionally, the current algorithm does not account for the moon distance and lunar phase, which can significantly impact observational constraints. A simple solution is to exclude tiles within a certain distance of the moon from the entire schedule, but this ignores movement of the moon through the night. Future versions of the scheduler could integrate these factors to optimize scheduling under realistic sky conditions.

These improvements would enhance the scheduler’s adaptability to real-time constraints, making it more effective for follow-up observations of GW events and other multi-messenger transients. With further refinements, \texttt{LUMOS} can play a critical role in maximizing the scientific yield of wide-field survey telescopes.

\section{Acknowledgments}

This work used Delta at the National Center for Supercomputing Applications (NCSA) through allocation AST200029, ``Towards a complete catalog of variable sources to support efficient searches for compact binary mergers and their products,'' from the Advanced Cyberinfrastructure Coordination Ecosystem: Services \& Support (ACCESS) program, which is supported by NSF grants
\#2138259, \#2138286, \#2138307, \#2137603, and \#2138296. We thank IBM for providing an academic license for CPLEX optimization software.

\bibliography{sample631}{}
\bibliographystyle{aasjournal}
\end{document}